\documentclass[12pt,preprint]{aastex}
\newcommand\Msun {M_{\odot}\ }

\def\etal{\rm et al.~}
\def\eg{\rm e.g., }
\def\ie{\rm i.e., }

\begin{document}

\title{Two distinct ancient components in the Sculptor Dwarf
Spheroidal Galaxy: First Results from DART\altaffilmark{1}}

\author{Eline Tolstoy\altaffilmark{2}, M.J. Irwin\altaffilmark{3}, 
A. Helmi\altaffilmark{2}, G. Battaglia\altaffilmark{2}, 
P. Jablonka\altaffilmark{4}, 
V. Hill\altaffilmark{4}, K.A. Venn\altaffilmark{5}, 
M.D. Shetrone\altaffilmark{6}, B. Letarte\altaffilmark{2}, 
A.A. Cole\altaffilmark{2},
F. Primas\altaffilmark{7}, P. Francois\altaffilmark{4}, 
N. Arimoto\altaffilmark{8}, K. Sadakane\altaffilmark{8},
A. Kaufer\altaffilmark{9}, T. Szeifert\altaffilmark{9}, 
T. Abel\altaffilmark{10}, 
}

\altaffiltext{1}{Based on FLAMES and UVES observations 
collected at the European Southern Observatory, 
proposals 71.B-0641 and 171.B-0588}

\altaffiltext{2}{Kapteyn Institute, University of Groningen, 
Postbus 800, 9700AV Groningen, the Netherlands}
\altaffiltext{3}{Institute of Astronomy, University of Cambridge, 
Madingley Road, Cambridge CB3 0HA, UK} 
\altaffiltext{4}{Observatoire de Paris, section de Meudon, 
5 Place Jules Janssen, F-92195 Meudon cedex, France}
\altaffiltext{5}{Macalester College, Saint Paul, MN 55105, USA}
\altaffiltext{6}{University of Texas, McDonald Observatory, USA}
\altaffiltext{7}{European Southern Observatory, 
Karl-Schwarzschild str 2, D-85748 Garching bei M\"{u}nchen, Germany}
\altaffiltext{8}{National Astronomical Observatory, 2-21-1 Osawa, 
Mitaka, Tokyo 181-8588, Japan}
\altaffiltext{9}{European Southern Observatory, Alonso de Cordova 3107, 
Santiago, Chile}
\altaffiltext{10}{Penn State University, State College, PA, USA}

\begin{abstract}

We have found evidence for the presence of two distinct ancient
stellar components (both $\geq$ 10~Gyr old) in the Sculptor dwarf
spheroidal galaxy.  We used the ESO Wide Field Imager (WFI) in
conjunction with the VLT/FLAMES spectrograph to study the properties
of the resolved stellar population of Sculptor out to and beyond the tidal
radius. We find that two components are discernible in the spatial
distribution of Horizontal Branch stars in our imaging, and in the
[Fe/H] and v$_{hel}$ distributions for our large sample of
spectroscopic measurements. They can be generally described as a
``metal-poor'' component ([Fe/H]$< -1.7$) and a ``metal-rich''
component ([Fe/H]$> -1.7$).  The metal-poor stars are more spatially
extended than the metal-rich stars, and they also appear to be 
kinematically distinct.  These results provide an important insight
into the formation processes of small systems in the early universe
and the conditions found there.  Even this simplest of galaxies
appears to have had a surprisingly complex early evolution.

\end{abstract}


\section{Introduction}

The Sculptor (Scl) Dwarf Spheroidal galaxy (dSph) is a close companion
of the Milky Way at high galactic latitude ({\it b}~=~$-$83$^o$).
Located at a distance of 72 $\pm 5$~kpc (Kunkel \& Demers 1977), with
v$_{hel} =$ 109.9 $\pm 1.4$ km/s (Queloz, Dubath \& Pasquini 1995) it
has a low total (dynamical) mass, 1.4 $ \pm 0.6 \times 10^7 \Msun$
(Queloz \etal 1995).  Irwin \& Hatzidimitriou (1995) made the first
modern comprehensive determination of the physical properties of Scl.
They determined a core radius, r$_c = 5.8 \pm 1.6$ arcmin, and a tidal
radius, r$_t = 76 \pm 5$ arcmin with a position angle of 99 $\pm$ 1
degrees, and confirmed it to be a moderately luminous galaxy in the
dSph class (M$_V = -10.7 \pm 0.5$), with a modest central surface
brightness ($\Sigma_{o,V} = 23.5 \pm 0.5$ mag/arcsec$^2$).  Like most
dSph galaxies Scl does not appear to have much, if any, HI gas
(\eg Bouchard,
Carignan \& Mashchenko 2003).

Observations of Scl have revealed a sizeable population of RR
Lyrae variable stars (\eg Kaluzny \etal 1995), clearly indicating that
its stellar population contains a globular cluster age component.  Scl
also shows tantalizing evidence in its extended horizontal branch (HB)
for an unusual enrichment history (\eg Hurley-Keller, Mateo \& Grebel
1999; Majewski \etal 1999).  
Monkiewicz \etal (1999) have detected stars several magnitudes
below the oldest possible main-sequence turn-offs. Their accurate
photometry allowed them to conclude that the mean age of Scl is
similar to that of a globular cluster, but that there is an age spread
within this epoch indicative of an extended star formation history.
Thus the entire star formation history of Scl apparently lasted only a
few Gyr, and after this initial period of activity more than 10~Gyr
ago Scl appears to have been dormant.

There have been several previous spectroscopic studies of individual
stars in Scl, firstly to determine the kinematic 
properties (\eg Armandroff \& Da Costa 1986; Aaronson \& Olszewski
1987; Queloz et al. 1995) which have indicated a moderate
mass-to-light ratio M/L$= 9 \pm 6 $(M/L$_V$)$_\odot$.  This value is
significantly higher than that typical for a globular cluster, but
much lower than found for many dSphs (\eg Draco and Ursa Minor).
There have also been studies of the abundances of small samples of
individual Red Giant Branch (RGB) stars at low and high resolution
(Shetrone \etal 2003; Tolstoy \etal 2001, 2003).  These stars were
selected from the central region of Scl, and although they showed a
wider range in [Fe/H] than would be naively expected based on
colour-magnitude diagram (CMD) analysis, no conclusions could be drawn
on the large scale spatial variations of these properties.

Here we present the first results from the DART (Dwarf Abundances and Radial
velocities Team) large programme at ESO.  We present the results for
v$_{hel}$ and [Fe/H] measurements from our FLAMES spectroscopy of
401 RGB stars for the first galaxy in our sample, Scl dSph.
The relatively high signal/noise, S/N ($\approx$ 10-20 per pixel) of our
data has enabled us to derive both accurate metallicites ($\approx$
0.1 dex from internal errors) and radial velocities (to $\approx \pm$2
km/s).  This is the first time that a large sample of both types of
measurement has been made in a single galaxy. A more detailed
description of our data and data reduction techniques will be provided
in subsequent papers (Irwin \etal in prep; Hill \etal in prep;
Battaglia \etal in prep). Here we provide an overview of our first
intriguing results.

\section{Observations}

\subsection{Imaging}

The ESO Wide Field Imager (WFI) observations were obtained from the
ESO archive for the central region of Scl, and more extended
observations of outer fields were made by us.  We used the pipeline
reduction software developed by the Cambridge Astronomical Survey Unit
for processing mosaic camera imaging (Irwin \& Lewis 2001; Irwin \etal
2004).  These observations were used to study the photometric
properties of Scl stars out to the tidal radius and as target
selection for VLT/FLAMES observations of stars along the RGB.

\subsection{Spectroscopy}

All VLT/FLAMES observations were made in Medusa mode, giving $\sim$120
fibres to be placed over a 25 arcmin diameter field of view.  Most of
the observations were taken with the Low Resolution grating LR8
(which includes the Ca~II triplet (CaT) lines), apart from the central
pointing where High Resolution gratings were used (see Hill \etal 
in prep).  The CaT method is well calibrated to provide an accurate
estimate of the [Fe/H] abundance of an RGB star (Rutledge, Hesser \&
Stetson 1997; Cole \etal 2004) and also to provide an accurate radial
velocity measurement.  The FLAMES fields were placed at varying
distances from the centre of the galaxy to beyond the tidal radius.

The FLAMES data were all reduced, extracted and wavelength calibrated
using the GIRBLDRS pipeline\footnote{available at SouceForge,
http://girbldrs.sourceforge.net/} provided by the FLAMES consortium
(Geneva Observatory, Blecha \etal 2003).  For sky subtraction,
velocity and equivalent width (EW) estimation, we developed our own
software which was thoroughly checked on
multiple observations of the same fields taken at different times.
Wavelength shifts were generally found to be negligible ($\approx \pm$1
km/s).  Each target spectrum was automatically continuum-corrected and
cross-correlated with a zero-continuum Gaussian model CaT template.
With a velocity determined it is then straightforward to estimate the
EW of the individual CaT lines. We found that for reliable [Fe/H]
determinations from CaT calibrations a S/N $\geq$ 10 is required, and
for reliable v$_{hel}$ a S/N $\geq$ 7.5 is sufficient. All the plots
in this paper use the S/N $> 10$ data set, as we always consider both
the [Fe/H] and the v$_{hel}$ measurements.

We were able to verify the zero point accuracy for both velocity and
abundance determinations by comparison with independent
previous measurements of stars in Scl, which we had deliberately
reobserved.  We also have overlap between our own HR
and LR samples to verify consistency of velocity and [Fe/H]
measurements using different gratings and based on different abundance
determination methods.

\section{Analysis}

\subsection{Spatial Distribution}

Previous studies have already suggested that the spatial distribution
of the Horizontal Branch stars in Scl shows signs of a population
gradient, with the red horizontal branch stars (RHB) being more
tightly concentrated than the blue horizontal branch (BHB) stars (\eg
Hurley-Keller \etal 1999; Majewski \etal 1999).  Unlike previous wide
field studies, our WFI imaging data extends beyond the nominal tidal
radius and with an average 5-$\sigma$ limiting magnitude of V$=$23.5
and I$=$22.5 also probes well below the horizontal branch.  This has
enabled us to unequivocally demonstrate that the BHB and RHB stars
have markedly different spatial distributions (see Figure 1).

These RHB and BHB populations were selected from the WFI CMD shown in
Figure~2 using the selection boxes marked.  It is clear that the BHB
region is almost completely uncontaminated by foreground stars,
whereas the RHB zone includes a significant underlying foreground
component. To make a quantitative estimate of the contamination of the
RHB sample by foreground stars we selected a region of similar CMD
foreground occupancy, well away from the Scl stellar locus (also
outlined in Figure~2).  The radial density distributions of the RHB
component were then corrected by the expected foreground contamination
and the results for both the RHB and BHB distributions together with
their density ratio as a function of radial distance, are shown in the
lower panels of Figure 1.  The different spatial occupancy of the two
populations is striking and provides strong evidence that we are
seeing two distinct components.  The characteristics of the BHB and
RHB are also consistent with different ages (\eg $\leq$2~Gyr), or
different metallicities ($\Delta$[Fe/H]$\sim$0.7 dex), from
theoretical modelling of globular cluster horizontal branches by Lee
et al. (2001).

\subsection{Metallicity Distribution}

Our VLT/FLAMES spectroscopic sample provides measurements of
[Fe/H] for 401 individual candidate RGB-selected stars with final
S/N$>$10, of which 308 have a high membership probability
(these are plotted in Figure~3), defined to be 80 km/s $<$
v$_{hel}$ $<$ 150 km/s \ie approximately within 3$\sigma$ of
v$_{sys}$ (see Figure 4).  At the high Galactic latitude of Scl
v$_{hel}$ is a good estimator of membership probability assuming
v$_{sys} = 110$ km/s, and $\sigma \approx$ 10 km/s.

For those RGB stars which were determined to have a high probability
of membership, we show the distribution of [Fe/H] measurements as a
function of elliptical radius (the equivalent distance along the
semi-major axis) from the centre of Scl in the upper panel of
Figure~3.  A well-defined metallicity gradient is apparent with a
similar scale size to the RHB versus BHB spatial distributions.  The
lower panel of this figure highlights the abrupt change in the [Fe/H]
distribution that occurs at about r$= 0.2$~deg (2r$_c$) and also the
broad range in [Fe/H] over the whole galaxy.  Although both inner and
outer distributions cover a broad range of values, the inner is
centred at [Fe/H]$= - 1.4$ with a range of $\pm 0.6$dex and the outer
distribution is centred at [Fe/H] $= -2$ with a similar range of $\pm
0.6$dex.  The drop in $<$[Fe/H]$>$ at 2r$_c$ from the centre also
corresponds to the transition between the dominance of the RHB and BHB
populations seen in Figure 1. This is not seen if the colour of the
RGB alone is taken as an indication of metallicity.

\subsection{Kinematics}

In the upper panel of Figure~4 we plot the sample of v$_{hel}$
measurements versus elliptical radius for which the continuum S/N $>
10$.  The distinction between foreground stars and members of Scl is
generally unambiguous.  The clustering around the systemic velocity of
v$_{sys} = 110$ km/s is clear and is highlighted by the dashed line
and the dotted 3$\sigma$ deviation limits.  Although present,
foreground contamination can be seen to be almost
negligible, even in the outer regions, which simplifies subsequent
analysis.

We have divided our ``member'' sample into metal-poor ([Fe/H$< -1.7$)
and metal-rich ([Fe/H]$>-1.7$) distributions. This difference is
quantified in the lower panels of Figure~4 which shows the v$_{hel}$
distributions over three radial zones for the different metallicity
components separately.  The variation in v$_{sys}$ and $\sigma$ in
these three regions is computed using the maximum likelihood approach
described in Hargreaves \etal (1994), but modified to directly compute
the likelihood function over a 2D search grid.  We found that
v$_{sys}$ does not vary significantly between different 
populations or regions, but $\sigma$ does (see
Figure 4).

One obvious feature of the velocity distributions is the difference in
the kinematic properties of the two metallicity components.  In
general the more metal poor component has a higher velocity dispersion
than the more metal rich component. This difference is most striking
in the central panel ($0.2^\circ <$ r $< 0.5^\circ$) of the triptych
where $\sigma$ is 11.0$\pm$1.0 km/s and 6.8$\pm$0.8 km/s 
respectively.  The swap
in the dominance of the components is stark in the outer zone (r $>$
0.5), where there are hardly any stars at all in the metal-rich
category and provides compelling evidence that we are indeed
observing two dynamically different components in Scl.

\section{Conclusions}

Our FLAMES results show that Scl contains two distinct stellar
components, one metal-rich, $-0.9 >$ [Fe/H] $> -1.7$, and one
metal-poor, $-1.7 >$ [Fe/H] $> -2.8$. The metal-rich population is
more centrally concentrated than the metal poor, and on average
appears to have a lower velocity dispersion, $\sigma_{metal-rich} = 7
\pm 1$ km/s, whereas $\sigma_{metal-poor} = 11 \pm 1$ km/s 
(see Figure~4).  Our
WFI results show that there is a comparable difference over the same
scale in the spatial distribution of BHB stars and 
RHB stars.  Plausibly combining these results enables us 
to distinguish these
two distinct components on the basis of kinematics (Figure 4),
metallicity (Figure 3) and spatial distribution (Figure 1).
These large samples allow us for
the first time to quantify these large scale trends in a statistically
meaningful fashion.

There are indications that this is a common feature of dSph galaxies.
Our preliminary analysis of v$_{hel}$ and [Fe/H]
measurements in the other galaxies in our sample (Fornax and Sextans
dSph, Battaglia et al., in prep) also shows very similar characteristics to
Scl, especially in the most metal poor component.  The mix of
populations between the inner and outer regions is clearly changing.
The radial velocity studies of Wilkinson \etal (2004) and Kleyna \etal 
(2004) have also considered the possibility that kinematically distinct
components exist in Ursa Minor, Draco and Sextans.

What mechanism could create two ancient stellar components in a small
dwarf spheroidal galaxy?  A simple possibility is that
the formation of these dSph galaxies began with an initial burst
of star formation, resulting in a stellar population with a mean
[Fe/H] $\leq - 2$.  Subsequent supernovae explosions from this initial
episode could have been sufficient to cause gas (and
metal) loss such that star formation was inhibited until the remaining
gas could sink deeper into the centre and begin star formation again
(\eg Mori, Ferrara \& Madau 2002).  Thus the subsequent generation(s)
of stars would 
inhabit a region closer to the centre of the galaxy, and have a
higher average metallicity ($\geq -1.4$) and different kinematics. 
Another possible cause is external influences, such as
minor mergers, or accretion of additional gas.  It might also be
that events surrounding the Epoch of Reionisation, strongly influenced the
evolution of these small galaxies (\eg Bullock, Kravtsov \& Weinberg
2001) and resulted in the stripping or photoevaporation of the outer
layers of gas in the dSph, meaning that subsequent more metal enhanced
star formation occured only in the central regions.

The full abundance analysis of the FLAMES HR data (Hill et al. in
prep) will provide more details of the chemical enrichment history of
Scl. This will hopefully enable us to distinguish between two
episodes of star formation, each of which created the separate
kinematic components seen here, or more continuous star formation,
manifested as a gradient in velocity dispersion and
metallicity from the centre of the galaxy.

Here we have shown that even dSphs, the smallest galactic systems, did
not form simply, as might be assumed.  The wealth of data presented
here points to the presence of dynamically distinct components
in these galactic building blocks, or templates, and
provides a stimulus to theories of the formation and evolution of
dSphs.

\acknowledgments

ET gratefully acknowledges support from a fellowship of the Royal
Netherlands Academy of Arts and Sciences, and thanks Renzo Sancisi
for a careful reading of the draft.
KAV thanks the NSF for support through a CAREER award, AST-9984073.
MDS would like to thank the NSF for partial support under AST-0306884.

\begin{figure}[!ht]
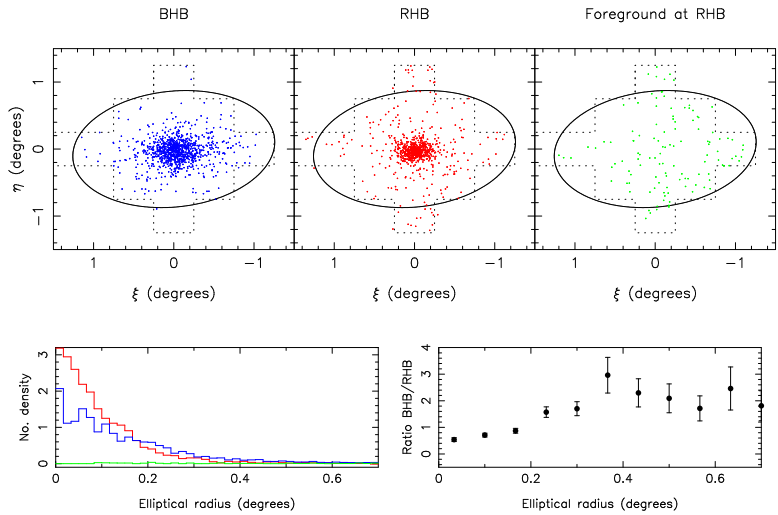

{\centering \scalebox{0.6}{
\includegraphics[angle=270,scale=.75]{fig1a.eps}} \par}
\vskip 0.5cm
{\centering \scalebox{0.6}{
\includegraphics[angle=270,scale=0.67]{fig1b.eps}} \par}

\caption{
The distribution of Horizontal Branch stars from WFI
imaging of the Scl dSph.  The upper panels show the different spatial
distributions of blue horizontal branch stars (BHB), and red
horizontal branch stars (RHB) as selected from the M$_{v}$, V$-$I
Colour-Magnitude Diagram (CMD) shown in Figure 2.  Also shown, to
illustrate the foreground contamination in the RHB distribution, are a
CMD-selected sample of foreground stars to match the RHB contamination
density (see Figure 2).  The ellipse is the tidal radius of Scl as
defined by Irwin \& Hatzidimitriou (1995).  The lower panels
quantitatively show the spatial variation of number density per unit
area of these three population components (BHB in blue, RHB in red and
foreground in green) and the ratio of BHB to RHB number densities
after correcting the RHB component for foreground contamination. The
physical scale at the distance of Sculptor is 1 degree to 1.4~kpc.
}
\end{figure}

\begin{figure}[!ht]
{\centering \scalebox{0.6}{
\includegraphics[scale=.8]{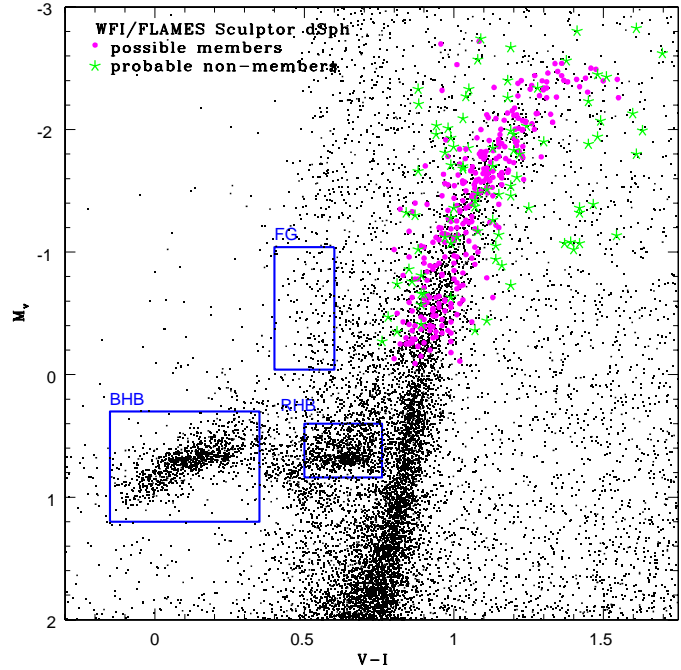}} \par}
\caption{
The Colour-Magnitude Diagram 
for the WFI coverage of Scl shown in Figure~1.  The green and
magenta symbols are the Red Giant Branch stars which were observed
with VLT/FLAMES and for which we have accurate v$_{hel}$ and [Fe/H]
measurements (S/N$>$10).  The potential members of Scl are shown as
magenta dots and non-members shown as green stars.  Also shown are the
regions used to define the BHB, RHB and foreground comparison 
sample (FG).
}
\end{figure}

\begin{figure}[!ht]
{\centering \scalebox{0.6}{
\includegraphics[scale=0.87]{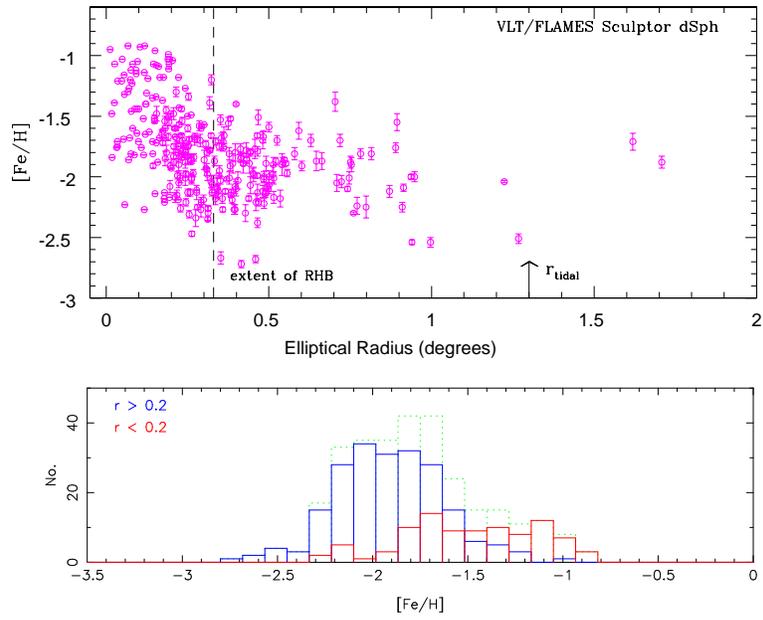}} \par}
\vskip -5.5cm
{\centering \scalebox{0.6}{
\includegraphics[angle=270,scale=.65]{fig3b.eps}} \par}
\caption{ 
VLT/FLAMES spectroscopic measurements of [Fe/H] for the 308 potential RGB 
velocity members of Scl versus radius 
(top panel).  The lower panel shows the distribution of [Fe/H] for: the 
entire sample of Scl member RGB stars (dotted line); the 97 stars 
within the central, r$=$0.2 degree region (red solid line); and the 211
stars beyond r$>$0.2 degrees (blue solid line).
}
\end{figure}

\begin{figure}[!ht]
\vskip -2.5cm
{\centering \scalebox{0.6}{
\includegraphics[scale=0.87]{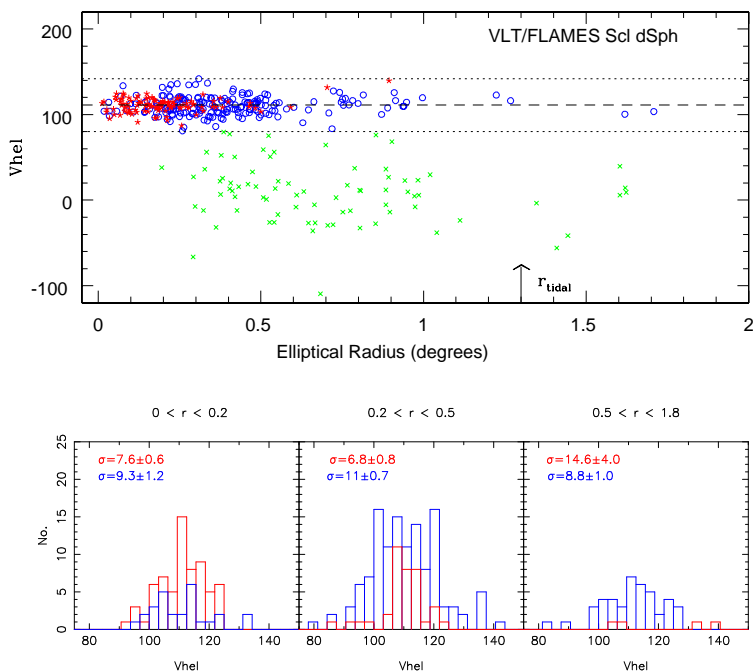}} \par}
\vskip -0.3cm
{\centering \scalebox{0.6}{
\includegraphics[angle=270,scale=.7]{fig4b.eps}} \par}
\caption{
The top panel shows v$_{hel}$ as a function of elliptical radius for
all stars satisfying S/N $>$ 10.  The
308 stars which are potential members are plotted as red stars
([Fe/H] $>$ -1.7) and blue circles ([Fe/H] $<$ -1.7), while the 93 
green crosses are assumed to be non-members.  The
lower panels quantify how the velocity distribution changes as a
function of metallicity and radius. The 
low metallicity distribution is plotted 
in blue, and the higher metallicity in red, as in the upper panel.
}
\end{figure}

\end{document}